\documentclass[preprint,showpacs,preprintnumbers,amsmath,amssymb,superscriptaddress]{revtex4}
\usepackage{graphicx}
\usepackage{dcolumn}
\usepackage{bm}
\usepackage{color}

\font\scripti=cmmi7
\font\scriptscripti=cmmi5
\def\sib#1{\setbox0 = \hbox{\scripti #1}
  \kern-.02em\copy0\kern-\wd0
  \kern.04em\box0} 
\def\ssib#1{\setbox0 = \hbox{\scriptscripti #1}
  \kern-.02em\copy0\kern-\wd0
  \kern.04em\box0} 
\font\tenib=cmmib10 
\skewchar\tenib='177 \skewchar\tenib='177 \skewchar\tenib='177
\def\pbold#1{\setbox0 = \hbox{$ #1 $}
  \kern-.022em\copy0\kern-\wd0
  \kern.011em\copy0\kern-\wd0
  \kern.011em\copy0\kern-\wd0
  \kern.011em\copy0\kern-\wd0
  \kern.011em\box0} 

\usepackage{graphicx}
\usepackage{dcolumn}
\usepackage{bm}

\def\lesssim{\ \raise.3ex\hbox{$<$}\kern-0.8em\lower.7ex\hbox{$\sim$}\ }
\def\gesim{\ \raise.3ex\hbox{$>$}\kern-0.8em\lower.7ex\hbox{$\sim$}\ }
\def\eave{\langle \mathcal{E}\rangle}
\def\eavehf{\langle \mathcal{E}\rangle_{\rm HF}}

\def\jpsj#1#2#3{J. Phys. Soc. Jpn. {\bf #1}, #2 (#3)}
\def\npa#1#2#3{Nucl. Phys. A {\bf #1}, #2 (#3)}
\def\npb#1#2#3{Nucl. Phys. B {\bf #1}, #2 (#3)}

\def\pra#1#2#3{Phys. Rev. A {\bf #1}, #2 (#3)}
\def\prb#1#2#3{Phys. Rev. B {\bf #1}, #2 (#3)}

\def\prl#1#2#3{Phys. Rev. Lett. {\bf #1}, #2 (#3)}

\def\rmp#1#2#3{Rev. Mod. Phys. {\bf #1}, #2 (#3)}

\newcommand{\blue}[1]{{\color{blue}{#1}}}

\begin{document}
\preprint{RIKEN-QHP-440}
\title{Goldstino spectrum
 in an ultracold Bose-Fermi mixture \\with explicitly broken supersymmetry}
\author{Hiroyuki Tajima}
\affiliation{Department of Mathematics and Physics, Kochi University, Kochi 780-8520, Japan}
\affiliation{RIKEN Nishina Center, Wako, Saitama 351-0198, Japan} 
\author{Yoshimasa Hidaka}
\affiliation{RIKEN Nishina Center, Wako, Saitama 351-0198, Japan} 
\affiliation{RIKEN iTHEMS, Wako,
  Saitama 351-0198, Japan}
\author{Daisuke Satow}
\affiliation{Arithmer Inc., Minato-Ku, Tokyo 106-6040, Japan}

\date{\today}

\begin{abstract}
We theoretically investigate a supersymmetric collective mode called Goldstino in a Bose-Fermi mixture.
The explicit supersymmetry breaking, which is unavoidable in cold atom experiments, is considered.
We derive the Gell-Mann--Oakes-Renner (GOR) relation for the Goldstino, which gives the relation between  the energy gap at the zero momentum
and the explicit breaking term.
We also numerically evaluate the gap of Goldstino above the Bose-Einstein condensation temperature within the random phase approximation (RPA).
While the gap obtained from the GOR relation coincides with that in the RPA for the mass-balanced system,
there is a deviation from the GOR relation in the mass-imbalanced system. 
We point out the deviation becomes large when the Goldstino pole is close to the branch point, although it is parametrically a higher order with respect to the mass-imbalanced parameter.
To examine the existence of the goldstino pole in realistic cold atomic systems,
we show how the mass-imbalance effect appears in $^6$Li-$^7$Li, $^{40}$K-$^{41}$K, and $^{173}$Yb-$^{174}$Yb mixtures. 
Furthermore, we analyze the Goldstino spectral weight in a $^{173}$Yb-$^{174}$Yb mixture with realistic interactions and show a clear peak due to the Goldstino pole.
As a possibility to observe the Goldstino spectrum in cold atom experiments,
we discuss the effects of the Goldstino pole on the fermionic single-particle excitation as well as the relationship between the GOR relation and Tan's contact.
\end{abstract}
\maketitle
\section{Introduction}
The supersymmetry is a symmetry with respect to an interchange between bosons and fermions~\cite{Fayet,Nilles,Witten2}.
While the existence of supersymmetry is expected in the context of particle physics,
its evidence or any indications have not been observed in high-energy experiments yet~\cite{Feng}.
However, apart from whether the supersymmetric partners such as squark exist or not in our world,
it is really an interesting problem to explore the consequences of the supersymmetry using fermions and bosons that are well established in condensed matter physics.  
\par
An ultracold atomic gas is nowadays one of the most useful systems to investigate quantum many-body phenomena, due to its controllability of physical parameters such as interaction, density, temperature, and quantum statistical properties of atoms by using isotopes~\cite{Dalfovo,Giorgini,Bloch}. 
In particular, the Feshbach resonance~\cite{Chin} enables us to investigate this atomic system from the weak-coupling to the strong-coupling limit in a systematic manner. 
In this regard, the supersymmetric properties of this system have been extensively discussed theoretically~\cite{Snoek,Shi,Yu,Yu2,HHLai,Satow1,Satow2,Bradlyn}. 
{Recently, Bose-Fermi mixtures with a small mass-imbalance between bosons and fermions such as  $^6$Li-$^7$Li~\cite{ENS2014,ENS2017,Ikemachi}, $^{39}$K-$^{40}$K~\cite{Falke}, $^{40}$K-$^{41}$K~\cite{MIT}, $^{84}$Sr-$^{87}$Sr~\cite{Tey}, {$^{87}$Rb-$^{87}$Sr~\cite{Barbe}},  $^{161}$Dy-$^{162}$Dy~\cite{Lu}, and
 $^{173}$Yb-$^{174}$Yb~\cite{Fukuhara,Sugawa} mixtures
 has been experimentally realized.}
The boson-boson or boson-fermion interactions in some of the mixtures can be tuned due to the magnetic Feshbach resonance~\cite{ENS2017,MIT,Pollack,DErrico,Kishimoto,Barbe}.
In this sense, examining supersymmetry in such cold atomic systems is promising.
\par
A remarkable feature of supersymmetry in a Bose-Fermi mixture is the emergence of NG mode called Goldstino~\cite{Witten,Salam,Lebedev:1989rz,Kratzert:2003cr,Kratzert:2002gh}. 
While a usual NG mode propagates as a bosonic mode, the Goldstino behaves as a fermionic mode. 
Such a fermionic collective excitation has also been predicted in quantum electrodynamics as well as quantum chromodynamics (QCD)~\cite{Lebedev:1989ev, Hidaka, Satow:2013oya}. 
Observation of this collective mode is really important to see the supersymmetric properties in a Bose-Fermi mixture that are realized in a table-top experiment.
Since the Goldstino is a fermionic collective mode associated with the broken supersymmetry,
it becomes a gapless mode when the system possesses the exact supersymmetry.
However, the explicit supersymmetry breaking such as mass-imbalance between fermions and bosons is unavoidable in cold atom experiments.
In such a case, the Goldstino has a finite energy-gap associated with explicit breaking parameters.
If one can observe the gapped Goldstino and its spectral properties agree with the result of theoretical analysis, it should be  evidence for the existence of supersymmetry in these systems.
Indeed, the first example of NG bosons in particle physics was pions, which are also gapped modes due to the explicitly broken chiral symmetry associated with the current quark mass~\cite{Hatsuda}.
\par
In this work, we theoretically examine the energy gap of Goldstino in a Bose-Fermi mixture with explicitly broken supersymmetry.
We focus on a few candidates for nearly-supersymmetric Bose-Fermi mixtures,
namely, $^6$Li-$^7$Li, $^{40}$K-$^{41}$K, and $^{173}$Yb-$^{174}$Yb mixtures.
We determine the thermodynamic properties of weakly interaction mixtures within the Hartree-Fock mean-field approximation above the Bose-Einstein condensation (BEC) temperature.
By developing a gap formula for the Goldstino, which corresponds to the Gell-Mann--Oakes--Renner (GOR) relation in QCD~\cite{GOR}, 
based on the memory function formalism~\cite{Mazenko}, we show how the explicit supersymmetry-breaking terms affect the Goldstino gap in these systems.
By comparing it with the numerical results of the random phase approximation (RPA),
we clarify that the effects of the branch point are significant in the presence of the mass-imbalance between fermions and bosons.
Furthermore, we discuss how to observe the goldstino gap from the single-particle spectral function of a Fermi atom.
{While the previous work is done by two of the authors are dedicated on the two-dimensional system~\cite{Satow1} and the three-dimensional one in the BEC phase~\cite{Satow2} with ideal situations such as supersymmetric interactions, in this paper, we discuss the three-dimensional system with realistic physical parameters above the Bose-Einstein condensation temperature $T_{\rm BEC}$.
Furthermore, we consider the case with the mass-imbalance where the fermionic mass is slightly lighter than the bosonic one.
In such a case, the effects of the branch point are more important. We show that, even when the mass-imbalance is small, the peak of the Goldstino disappear and it is buried in the continuum spectrum if the interaction is too weak. 
}
{We also note that, there is a work which considered a mass-imbalanced Bose-Fermi system~\cite{Shi}, in which the Bose-Fermi mixture trapped on optical lattice was considered, and the species of atoms were not specified. 
These points are improved in this paper.}
\par
This paper is organized as follows:
In Sec.~\ref{sec2}, we introduce our model and the formulation for thermodynamic quantities and the Goldstino gap within GOR and RPA.
In Sec.~\ref{sec3}, we show our numerical results in RPA on the Goldstino gap in a few Bose-Fermi mixture systems, and discuss it.
Section~\ref{sec:fermionSpectrum} is devoted to discussion on how the Goldstino pole can affect the fermionic single-particle spectrum, in order to suggest the possibility for detecting Goldstino in experiments.
We summarize our studies in Sec.~\ref{sec4}.
{In Appendix~\ref{sec:MemoryFunction}, we show the detailed derivation of the GOR relation based on the memory function formalism.}
We calculate the Goldstino spectral function in the free limit in Appendix~\ref{AppB} to check the absence of numerical {artifacts}.

\section{Formalism}
\label{sec2}
\subsection{Model}
We consider a non-relativistic Bose-Fermi mixture described by the Hamiltonian,
\begin{align}
\label{eq1}
H&=\int d^3\bm{r}\psi_b^{\dag}(\bm{r})\left(-\frac{\nabla^2}{2m_b}-\mu_b\right)\psi_{b}(\bm{r})
+\int d^3\bm{r}\psi_f^{\dag}(\bm{r})\left(-\frac{\nabla^2}{2m_f}-\mu_f\right)\psi_{f}(\bm{r})\cr
&+\frac{U_{bb}}{2}\int d^3\bm{r}\psi_b^\dag(\bm{r})\psi_b^\dag(\bm{r})\psi_b(\bm{r})\psi_b(\bm{r})
+U_{bf}\int d^3\bm{r}\psi_b^\dag(\bm{r})\psi_b(\bm{r})\psi_f^\dag(\bm{r})\psi_f(\bm{r}),
\end{align}
where $\psi_{b(f)}$ is the field operator of a boson (fermion) with a mass $m_{b(f)}$ and the chemical potential $\mu_{b(f)}$.
$U_{bb(bf)}$ is the coupling constant of a boson-boson (boson-fermion) interaction, which is assumed to be a contact-type. 
{These coupling constants are related to the scattering length $a_{bb(bf)}$ as
$U_{bb}=(4\pi a_{bb})/m_b$ and $U_{bf}=(2\pi a_{bf})/{m_r}$, respectively, where $m_r=1/(1/m_f+1/m_{b})$ is the reduced mass.}
{In this paper, we measure the interaction strength by using a dimensionless parameters $k_ba_{bb}$ and $k_ba_{bf}$, where $k_{b}=(6\pi^2N_b)^{{1}/{3}}$ is a momentum scale for boson density $N_{b}$.}
{In general, there is a non $s$-wave fermion-fermion interaction such as dipole-dipole interaction given by
\begin{align}
\label{eqVff}
V_{ff}=\frac{1}{2}\int d^3\bm{r}\psi_f^\dag(\bm{r})\psi_f^\dag(\bm{r}')U_{ff}(\bm{r}-\bm{r}')\psi_f(\bm{r}')\psi_f(\bm{r}).
\end{align}
Although it is negligible in several Fermi atoms such as $^6$Li and $^{40}$K far away from higher partial-wave Feshbach resonances at low temperature,
it would become significant in a $^{161}$Dy-$^{162}$Dy mixture with the large magnetic dipole moments~\cite{Lu}.
In this work, we consider the case in which  the dipole-dipole interaction is negligible for simplicity.
}
{We note that  the inter-component interaction $U_{bf}$ involves a factor $2$, in contrast to the intra-component interaction $U_{bb}$~\cite{FW}.}
When $m_{f}=m_{b}$, $\mu_{b}=\mu_{f}$, $U_{bb}=U_{bf}$, there is a supersymmetry corresponding to interchange between bosons and fermions:
$\psi_{b}\to \psi_{f}$ and $\psi_{f}\to \psi_{b}$. The corresponding Noether charges are
\begin{equation}
\label{eq2}
Q=\int d^3\bm{r}q(\bm{r}),\qquad
Q^{\dag}=\int d^3\bm{r}q^{\dag}(\bm{r}),
\end{equation}
which commute with the Hamiltonian, $[H, Q]=[H, Q^{\dag}]=0$.
Here, $q(\bm{r})=\psi_{f}(\bm{r})\psi_{b}^\dag(\bm{r})$ is the local operator that creates the boson and annihilates the fermion~\cite{Yu}.
Unlike the supersymmetry in relativistic systems, the anti-commutation relation between supercharges is not the Hamiltonian but the total particle number operator:
\begin{equation}
\{Q,Q^{\dag}\}
= 
\int d^3\bm{r}\psi_{f}^{\dag}(\bm{r})\psi_{f}(\bm{r})+\int d^3\bm{r}\psi_{b}^{\dag}(\bm{r})\psi_{b}(\bm{r}).
\end{equation}
In this sense, the supersymmetry in a non-relativistic Bose-Fermi mixture is a different type from that in relativistic theories.
The order parameter of supersymmetry breaking  is the total number density, $\langle\{Q, q^{\dag}(\bm{r})\} \rangle=\langle\psi_{b}^{\dag}(\bm{r})\psi_{b}(\bm{r})\rangle+\langle\psi_{f}^{\dag}(\bm{r})\psi_{f}(\bm{r})\rangle$,
which is always broken in a finite density system.
{For a spontaneous breaking of bosonic continuous symmetry, if the order parameter is expressed as the expectation value of the commutation relation between a charge and a charge density, the breaking pattern is called the type-B~\cite{Watanabe:2011ec,Watanabe:2012hr,Hidaka:2012ym,Watanabe:2014fva,Hayata:2014yga,Takahashi:2014vua}.
On the other hand, if no such an order parameter exists, the breaking pattern is called the type-A. The NG modes corresponding to the type-B typically exhibit the quadratic dispersion. A typical example of type-B NG mode is the magnon in a ferromagnet, in which the order parameter is expressed as the expectation value of the commutation relation between spins. 
Replacing the commutator by the anticommutator, we can identify the supersymmetry breaking pattern as the type-B.}
As in an ordinary symmetry breaking, the supersymmetry breaking leads to a gapless excitation.
If the excitation can be identified as a single-mode excitation, it is called the Goldstino. In general, the excitation may be located at a branch point where two or multi-particles continuum starts. This is especially the case for the non-interacting system, where there is no Goldstino. The excitation is the particle-hole one.
The interaction plays an important role in the existence of the Goldstino.
In the following analysis, we assume the existence of  the Goldstino excitation, and we numerically check it in the RPA in Sec.~\ref{sec3}.
Since the order parameter is expressed as the expectation value of the anti-commutation relation of the supercharge and its density, the Goldstino belongs to 
the type-B mode~\cite{Watanabe:2011ec,Watanabe:2012hr,Hidaka:2012ym,Watanabe:2014fva,Hayata:2014yga,Takahashi:2014vua}, which typically has a quadratic dispersion. 

In a realistic situation, the supersymmetry is explicitly broken because all parameters cannot be exactly tuned in experiments. 
The effect of the explicit breaking can be expressed as the commutation relation between the Hamiltonian and the supercharge:
\begin{equation}
\label{eq3}
[H,Q]=\int d^3\bm{r}\psi_b^\dag(\bm{r})\left(\chi\frac{\nabla^2}{2m_{r}}+\Delta\mu\right)\psi_f(\bm{r})
-\Delta U\int d^3\bm{r}\psi_b^\dag(\bm{r})\psi_b^\dag(\bm{r})\psi_b(\bm{r})\psi_f(\bm{r}),
\end{equation}
{where we define
\begin{align}
\Delta\mu&\equiv\mu_f-\mu_b,\\
\chi&\equiv\Bigl(\frac{1}{m_f}-\frac{1}{m_{b}}\Bigr)m_r= \frac{m_b-m_f}{m_b+m_f},\\
\Delta U&\equiv U_{bf}-U_{bb}.
\end{align}
}
These explicit breakings cause a finite gap of the Goldstino, whose formula is shown in the next subsection.
\par

\subsection{Gell-Mann--Oakes--Renner relation}
Pions are the Nambu-Goldstone (NG) bosons associated with the spontaneous breaking of chiral symmetry in QCD.
The Gell-Mann--Oakes--Renner (GOR) formula relates the pion mass and the current quark mass that explicitly breaks chiral symmetry~\cite{GOR}. 
We can generalize the GOR relation to that of Goldstino in a Bose-Fermi mixture.
For this purpose, we employ the memory function formalism~\cite{Mazenko},
which is a different formalism {from the one used} in the original derivation~\cite{GOR}. 
{
The derivation is slightly technical, so that we, here, show the only result. 
For readers who are interested in the derivation, see the Appendix~\ref{sec:MemoryFunction}.

We consider the retarded Goldstino propagator defined as
\begin{equation}
\label{eq4}
\Gamma^R(\bm{r},t) \equiv i\theta(t)\langle\{q(\bm{r},t),q^\dag(\bm{0},0) \}\rangle.
\end{equation}
After Fourier transformation, we obtain
\begin{equation}
\label{eq5}
\Gamma^R(\bm{p},\omega)=i\int_{-\infty}^\infty dt\int d^3\bm{r}e^{i\omega t-i\bm{p}\cdot\bm{r}}\theta(t)\langle\{q(\bm{r},t),q^\dag(\bm{0},0) \}\rangle.
\end{equation}
The energy gap is obtained from the pole of $\Gamma^R(\bm{p},\omega)$ in the complex $\omega$ plane.
Since we are interested in the zero-momentum gap of Goldstino, hereafter we take $\bm{p}=\bm{0}$.
The memory function formalism systematically decompose $\Gamma^R(\omega)$ into the following form:
\begin{equation}
\begin{split}
\label{eq:retardedGreenFunction}
\Gamma^R(\omega)= \frac{-N}{\omega+\Omega +i\Phi(\omega)},
\end{split}
\end{equation}
where $N=\langle\psi_{f}^{\dag}(\bm{r})\psi_{f}(\bm{r})\rangle+\langle\psi_{b}^{\dag}(\bm{r})\psi_{b}(\bm{r})\rangle$
is the total number density. $\Phi(\omega)$ and  $ \Omega =  \langle \{ [H, Q],q^{\dag}(\bm{0},0)\}\rangle/N$ are called the dynamic and static parts of the memory function. We do not show the explicit form of $\Phi(\omega)$; the important point is $\Phi(\omega)$ is parametrically higher oder 
compared with $\Omega$ with respect to the explicit breaking term (See the Appendix~\ref{sec:MemoryFunction} for more details). Therefore, at the leading order of the explicit breaking term, the energy gap $\omega_{G}$ is expressed as
\begin{equation}
\label{eq12}
\omega_{G} =\omega_{G}^{\rm GOR}\equiv -\frac{1}{N}\langle\{[H,Q],q^\dag(\bm{0},0)\}\rangle.
\end{equation}
We emphasize that this formula works for any supersymmetric Hamiltonian with a small explicit breaking term and local interactions because we have not employed the specific form of the Hamiltonian. 
The gap is linearly proportional to the explicit breaking term, whose property can be understood as the type-B breaking~\cite{Watanabe:2011ec,Watanabe:2012hr,Hidaka:2012ym,Watanabe:2014fva,Hayata:2014yga,Takahashi:2014vua}. 
In contrast, the type-A breaking predicts that the gap is proportional to the square root of the explicit breaking term.
We note that although the dynamic part $\Phi(\omega)$ is higher order, it might not be small if there is a singularity in $\Phi(\omega)$. 
As is seen later,  this is the case when the branch point is close to $\omega_{G}^{\rm GOR}$.
}

For the Hamiltonian~\eqref{eq1} that we employ in the present paper,
using Eq.~(\ref{eq3}), we can obtain
\begin{align}
\label{eq13}
\{[H,Q],q^\dag(\bm{r},0)\}&=\frac{\chi}{2m_r}\left[\{\nabla^2\psi_{b}^\dag(\bm{r})\}\psi_b(\bm{r})+\psi_f^\dag(\bm{r})\nabla^2\psi_f(\bm{r})\right]\cr
&\quad+\Delta\mu\left[\psi_b^\dag(\bm{r})\psi_b(\bm{r})+\psi_f^\dag(\bm{r})\psi_f(\bm{r})\right]\cr
&\quad-\Delta U\left[\psi_b^\dag(\bm{r})\psi_b^\dag(\bm{r})\psi_b(\bm{r})\psi_b(\bm{r})
+2\psi_b^\dag(\bm{r})\psi_b(\bm{r})\psi_f^\dag(\bm{r})\psi_f(\bm{r})\right].
\end{align}
Therefore, the goldstino gap in the present model is given by
\begin{align}
\label{eq14}
\omega_{G}^{\rm GOR}&= -\Delta\mu-\frac{\chi}{2m_rN}\left[\langle\{\nabla^2\psi_{b}^\dag(\bm{r})\}\psi_b(\bm{r})+\psi_f^\dag(\bm{r})\nabla^2\psi_f(\bm{r})\rangle\right]\cr
&\quad+\frac{\Delta U}{N}\left[\langle\psi_b^\dag(\bm{r})\psi_b^\dag(\bm{r})\psi_b(\bm{r})\psi_b(\bm{r})\rangle\right.+\left.2\langle\psi_b^\dag(\bm{r})\psi_b(\bm{r})\psi_f^\dag(\bm{r})\psi_f(\bm{r})\rangle\right]\notag\\
&=
-\Delta\mu+\chi \eave
+\frac{2\Delta U}{U}\langle V\rangle,
\end{align}
where $U\equiv (U_{bf}+U_{bb})/2$.
Here $\eave$ and $\langle V\rangle $ are the average kinetic and {interaction} energy of one particle per volume:
\begin{align}
\label{eqEave}
\eave &= \frac{1}{N}\left\langle 
\psi_b^{\dag}(\bm{r})\frac{-\nabla^2}{2m_r}\psi_{b}(\bm{r})
+\psi_f^{\dag}(\bm{r})\frac{-\nabla^2}{2m_r}\psi_{f}(\bm{r})
\right\rangle,\\
\label{eqVave}
\langle V\rangle &=\frac{1}{N}\left\langle\frac{U}{2}\psi_b^\dag(\bm{r})\psi_b^\dag(\bm{r})\psi_b(\bm{r})\psi_b(\bm{r})+ U\psi_b^\dag(\bm{r})\psi_b(\bm{r})\psi_f^\dag(\bm{r})\psi_f(\bm{r})\right\rangle.
\end{align}
{Here, we take the limits of $\chi \rightarrow 0$ ($m_b\rightarrow m_f$) and $\Delta U\rightarrow 0$ ($U_{bb}\rightarrow U_{bf}$) to suppress the higher-order breaking terms being proportional to $\chi^2$ and $(\Delta U^2$ when we define $\eave$ and $\langle V \rangle$ in Eqs. (\ref{eqEave}) and (\ref{eqVave}).}
These parameters can be also expressed by the pressure $P(T,\mu, m_{f},m_{b}, U_{bb}, U_{bf})$ as a function of $T$, $\mu$, $m_{f}$, $m_{b}$, $U_{bb}$, and $U_{bf}$,
\begin{align}
\eave &= \frac{1}{m_r}\left(m_{f}^2\frac{\partial P}{\partial m_{f}}+m_{b}^2\frac{\partial P}{\partial m_{b}}\right),\\
\langle V\rangle &= -U\left(\frac{\partial P}{\partial U_{bf}}+\frac{\partial P}{\partial U_{bb}}\right).
\end{align}
We note that, this result is correct up to the first order in explicit symmetry breaking, and we did not use any approximations such as RPA in its derivation.
We also note that the expectation values of the local operator $\langle\psi_b^\dag(\bm{r})\psi_{b(f)}^\dag(\bm{r})\psi_b(\bm{r})\psi_{b(f)}(\bm{r})\rangle$ in zero-range models are known to be associated with the so-called Tan's contact $C_{bb(bf)}$ ~\cite{Tan1,Tan2,Tan3,Braaten} as
\begin{align}
\langle V\rangle&=\frac{U}{N}\left[\frac{C_{bf}}{(4\pi  a_{bf})^2}+\frac{C_{bb}}{(4\pi a_{bb})^2}\right].
\end{align}
The universal relations with respect to this quantity is expected to hold even in the weakly repulsive case~\cite{Diederix,Qu}.
Indeed, $C_{bb}$ is analytically obtained within the mean-field Bogoliubov theory at $T=0$ in Refs.~\cite{Diederix,Liu}.  
The GOR relation is therefore rewritten as
\begin{align}
\label{eqC}
\omega_{G}^{\rm GOR}&=-\Delta\mu+\chi\eave+2\frac{\Delta U}{N}\left[\frac{C_{bf}}{(4\pi  a_{bf})^2}+\frac{C_{bb}}{(4\pi a_{bb})^2}\right]\cr
&=-\Delta\mu+\chi\eave+\frac{1}{2\pi N}\left(\frac{a_{bf}}{{2}m_r}-\frac{a_{bb}}{m_b}\right)\left(\frac{C_{bf}}{a_{bf}^2}+\frac{C_{bb}}{a_{bb}^2}\right).
\end{align}
Since Tan's contact can precisely be observed,
this relation is also useful to address the Goldstino properties in recent experiments.
However, a strong {repulsive} interaction beyond the present {weak-coupling} mean-field approximation generally involves {an effective range correction acting as a momentum cutoff to avoid an ultraviolet divergence} in a three-dimensional system~\cite{Randeria,Palestini}.
{In this case, one has to extend Eq.~(\ref{eqC}) to the relation with the effective ranges of the interactions. }
In this paper, we restrict ourselves in the weak-coupling regime and it is left for {future work}.
Since we assume a homogeneous case with the translational symmetry, we can take $\bm{r}\rightarrow 0$ after acting $\nabla^2$ in the terms in $\langle \mathcal{E}\rangle$.
{We also note that the GOR relation derived in this paper is valid in both below and above the Bose-Einstein condensation temperature $T_{\rm BEC}$.
To address the BEC phase below $T_{\rm BEC}$, one has to take the mean-field term associated with the condensate into account~\cite{Andersen}.
In this paper, we consider the normal phase above $T_{\rm BEC}$ for simplicity.
}
\subsection{Mean-field approximation}
In this paper, we employ the weak-coupling mean-field approximation to calculate the Goldstino gap by using the GOR relation~(\ref{eq14}).
At a weak coupling, the thermal average with respect to the interaction term in Eq.~(\ref{eq14}) can be approximated as
\begin{align}
\label{eq15}
\langle\psi_b^\dag(\bm{r})\psi_b^\dag(\bm{r})\psi_b(\bm{r})\psi_b(\bm{r})\rangle\simeq2N_b^2,\\
\label{eq16}
\langle\psi_b^\dag(\bm{r})\psi_b(\bm{r})\psi_f^\dag(\bm{r})\psi_f(\bm{r})\rangle\simeq N_b N_f,
\end{align}
where the particle number densities $N_{b(f)}$ is obtained as
\begin{align}
\label{eq17}
N_b&=\int \frac{d^{3}\bm{q}}{(2\pi)^{3}} n_b(\xi_{\bm{q}}^b),\\
\label{eq18}
N_f&=\int \frac{d^{3}\bm{k}}{(2\pi)^{3}} n_f(\xi_{\bm{k}}^f),
\end{align}
where $n_{b(f)}(x)=1/(\exp({x/T})\mp 1)$ is the Bose (Fermi) distribution function.
Here we have defined $\xi_{\bm{q}}^b=q^2/(2m_b)-\mu_b+\Sigma_b^H$ and $\xi_{\bm{k}}^f=k^2/(2m_f)-\mu_f+\Sigma_f^H$. 
The Hartree shift $\Sigma_{b(f)}^H$ is given by
\begin{align}
\label{eq19}
\Sigma_{b}^{H}&=2U_{bb}N_b+U_{bf}N_f,\\
\label{eq20}
\Sigma_f^H&=U_{bf}N_b.
\end{align}
Substituting Eqs.~(\ref{eq17}) and (\ref{eq18}) into Eq.~(\ref{eq14}), one can obtain
\begin{equation}
\label{eq21}
\omega_G^{\rm GOR}=-\Delta\mu+\chi\eavehf+2N_b\Delta U.
\end{equation}
Here, we defined
\begin{equation}
\eavehf\equiv
\frac{1}{N}\int \frac{d^{3}\bm{k}}{(2\pi)^{3}}[n_b(\xi_{\bm{k}}^b)+n_f(\xi_{\bm{k}}^f)]\frac{k^2}{2m_r}.
\end{equation}
In particular, in the mass-balanced case ($m_b=m_f$) {relevant for a $^{87}$Sr-$^{87}$Rb mixture}, one can find
\begin{equation}
\label{eq22}
\omega_G^{\rm GOR}=-\Delta\mu+2N_b\Delta U.
\end{equation}
Our result agrees with the result in Ref.~\cite{HHLai} obtained in a tight-binding model.%
{We note that Eq.~(\ref{eq22}) obtained in the normal phase
 is different from the result in Ref.~\cite{Bradlyn}, which considered the BEC phase at zero temperature.}
We also note that in the mean-field approximation one can obtain Tan's contacts as
$C_{bf}=16\pi^2a_{bf}^2N_bN_f$ and $C_{bb}=16\pi^2a_{bb}^2N_b^2$. One can reproduce Eq.~(\ref{eq21}) by substituting them into Eq.~(\ref{eqC}).
\subsection{Random phase approximation}
We compare the results of the GOR relation with the RPA calculation to see effects of continuum and higher order correction in the explicit breaking term.
\begin{figure}[t]
\begin{center}
\includegraphics[width=7.5cm]{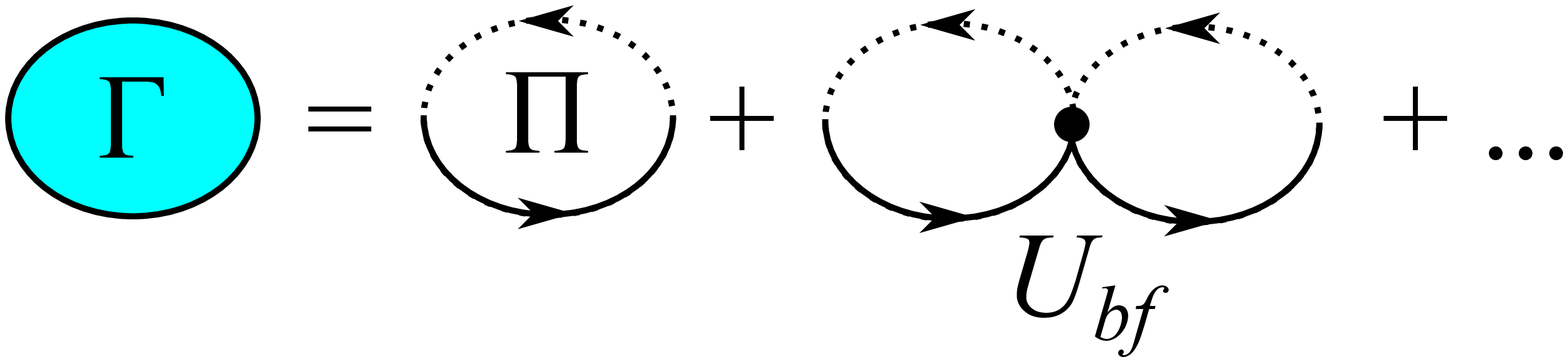}
\end{center}
\caption{Feynman diagrams for the Goldstino propagator $\Gamma$ consisting of RPA series of boson-fermion bubble $\Pi$.
The solid (dashed) line and
the black dot represent the fermion (boson) propagator and the boson-Fermion interaction $U_{bf}$, respectively.
}
\label{fig1}
\end{figure}
We consider the series of fermion-boson bubble $\Pi$ diagrammatically described by Fig.~\ref{fig1}.
The explicit form of the Goldstino propagator $\Gamma^R$ reads
\begin{equation}
\label{eq23}
\Gamma^{R}(\bm{p},\omega)=\frac{\Pi(\bm{p},\omega)}{1+U_{bf}\Pi(\bm{p},\omega)},
\end{equation}
where
\begin{equation}
\label{eq24}
\Pi(\bm{p},\omega)=-\int \frac{d^{3}\bm{k}}{(2\pi)^{3}}\frac{n_f(\xi_{\bm{k}}^f)+n_b(\xi_{\bm{k-p}}^b)}{\omega+i\delta+\xi_{\bm{k-p}}^b-\xi_{\bm{k}}^f}
\end{equation}
is a bubble diagram with respect to the fermion-boson exchange.
{{Here}} $\delta$ is an infinitesimally small value.
In the numerical calculations, $\delta$ is taken to be $10^{-3}\varepsilon_b$ in this paper.
{Here $\varepsilon_b=(6\pi^2N_b)^{2/3}/(2m_b)$ is the energy scale associated with the boson density $N_b$}.
{{We note that, the continuum is generated when the kinematics of $1$ to $2$ scattering is possible for multiple $\omega$ due to multiple $\bm{k}$:
\begin{equation}
\label{eq-kinematics}
\omega+\xi_{\bm{k-p}}^b=\xi_{\bm{k}}^f.
\end{equation}
}}
For $\bm{p}=\bm{0}$ and $\chi>0$, the branch point is located at $\omega_{\rm BP} = \xi_{\bm{0}}^f-\xi_{\bm{0}}^b=-\Delta\mu+2\Delta U N_{b}-U_{bf}N$.
There is the continuum spectrum for $\omega\geq \omega_{\rm BP}$. 
Since $\omega_{\rm BP}$ can be written as $\omega_{\rm BP}=\omega_G^{\rm GOR}-{\chi}\eavehf-U_{bf}N$,
$\omega=\omega_G^{\rm GOR}$ is always in the continuum for $\chi>0$.
In contrast, when $\chi<0$, the continuum spectrum exists for $\omega\leq\omega_{\rm BP}$.
Thus, $\omega=\omega_G^{\rm GOR}$ is in the continuum when $U_{bf}N\leq-\chi \eavehf$.

The Goldstino gap is obtained by the zero point of the denominator of $\Gamma^{R}(\bm{0},\omega)$, i.e, $1+U_{bf}\Pi(\bm{0},\omega)=0$.
In the mass-balanced case, one can analytically estimate the goldstino gap from
\begin{align}
0=1+U_{bf}\Pi(\bm{0},\omega_{G})&=1-U_{bf}{\int \frac{d^{3}\bm{k}}{(2\pi)^{3}}}\frac{n_f(\xi_{\bm{k}}^f)+n_b(\xi_{\bm{k}}^b)}{\omega_{G}+i\delta
+\Delta\mu-2\Delta UN_{b}+U_{bf}N}\notag\\
&=\frac{\omega_{G}+\Delta\mu-2\Delta UN_{b}}{\omega_{G}+i\delta
+\Delta\mu-2\Delta UN_{b}+U_{bf}N}
\end{align}
and therefore
\begin{equation}
\label{eq26}
\omega_G=-\Delta\mu+2N_b \Delta U.
\end{equation}
{{We note that, the $\bm{k}$ dependence completely vanishes from Eq.~(\ref{eq-kinematics}) at $\bm{p}=\bm{0}$, and therefore the width of the continuum becomes zero.}}
Equation~(\ref{eq26}) coincides with the GOR relation given by Eq.~(\ref{eq22})~\cite{Shi,Satow1,Satow2}.
Beyond the RPA, there will be corrections coming from interactions between quasi-particles.
Since the differences of chemical potentials and interactions simply induces a shift of the Goldstino pole, the supersymmetric collective mode can experimentally be confirmed {{by checking}} the interaction and chemical potential dependences of the gap in a weakly interacting mass-balanced mixture.

\par
On the other hand, in the presence of the mass-imbalance between bosons and fermions,
there is a correction to the GOR relation, which is parametrically higher order in the explicit breaking term.
However, the correction may not be small if the branch point is close to $\omega_G^{\rm GOR}$.
To see this, we parametrize the denominator of $\Gamma^{R}(\bm{0},\omega)$ as
\begin{equation}
1+U_{bf}\Pi(\bm{0},\omega)
=\frac{1}{U_{bf}N}\left[\omega-\omega_G^{\rm GOR}-\tilde{\Phi}(\omega) \right],
\end{equation}
where
\begin{equation}
\tilde{\Phi}(\omega)=
\frac{1}{N}{\int \frac{d^{3}\bm{k}}{(2\pi)^{3}}}\left[n_f(\xi_{\bm{k}}^f)+n_b(\xi_{\bm{k}}^b)\right]
\frac{\left[\omega+\Delta\mu-\chi k^{2}/(2m_{r})-2\Delta U N_{b}\right]^{2}}{
\omega+i\delta-\chi k^{2}/(2m_{r})-\omega_{\rm BP}}.
\end{equation}
$\tilde{\Phi}(\omega)$ plays a similar role of the dynamic part of the memory function defined in Eq.~\eqref{eq:Phiz},
although the definition is different.
At $\omega=\omega_G^{\rm GOR}$, $\tilde{\Phi}(\omega_G^{\rm GOR})$ is explicitly proportional to $\chi^{2}$:
\begin{equation}\label{eqphi}
\tilde{\Phi}(\omega_G^{\rm GOR})=
\chi^{2}\frac{1}{N}{\int \frac{d^{3}\bm{k}}{(2\pi)^{3}}}\left[n_f(\xi_{\bm{k}}^f)+n_b(\xi_{\bm{k}}^b)\right]
\frac{\left[k^{2}/(2m_{r})-{\eavehf}\right]^{2}}{
\omega_G^{\rm GOR}-\omega_{\rm BP}-\chi k^{2}/(2m_{r})+i\delta}.
\end{equation}
From this expression, the correction in the $\chi^{2}$ order  is evaluated as
\begin{equation}
\tilde{\Phi}(\omega_G^{\rm GOR})\simeq
\frac{\chi^{2}}{U_{bf}N}\frac{1}{N}{\int \frac{d^{3}\bm{k}}{(2\pi)^{3}}}\left[n_f(\xi_{\bm{k}}^f)+n_b(\xi_{\bm{k}}^b)\right]
\Bigl({\eavehf}-\frac{k^{2}}{2m_{r}}\Bigr)^{2}.
\end{equation}
We can estimate the scale of $\tilde{\Phi}(\omega_G^{\rm GOR})$ as $\chi^{2}\eavehf^{2}/ (U_{bf}N)$, where the integral of 
$({\eavehf}-{k^{2}}/{2m_{r}})^{2}$ is estimated  to be $\eavehf^{2}$. Since $\chi\eavehf\sim \omega_G^{\rm GOR}$ and $\omega_{\rm BP}\sim  U_{bf}N$ for a small explicit symmetry breaking case, 
we obtain
\begin{equation}
\begin{split}
\label{eq:orderEstimate}
\tilde{\Phi}(\omega_G^{\rm GOR})\sim \omega_G^{\rm GOR}\left| \frac{\omega_G^{\rm GOR}}{\omega_{\rm BP}}\right|.
\end{split}
\end{equation}
Similarly, we can estimate $n$-th order in $\chi$ as of order $\omega_G^{\rm GOR}| {\omega_G^{\rm GOR}}/{\omega_{\rm BP}}|^{n-1}$. 
This expansion breaks down if $| {\omega_G^{\rm GOR}}/{\omega_{\rm BP}}|$ is not small even though $\chi\ll 1$. 

When $\chi>0$, there is a contribution from the imaginary part  of $\tilde{\Phi}(\omega_G^{\rm GOR})$ to the dispersion relation,
which can be analytically evaluated as
\begin{align}
-\mathrm{Im}\tilde{\Phi}(\omega_G^{\rm GOR})&=
\chi^{2}\frac{\pi}{N}{\int \frac{d^{3}\bm{k}}{(2\pi)^{3}}}\left[n_f(\xi_{\bm{k}}^f)+n_b(\xi_{\bm{k}}^b)\right]
\Bigl(\frac{k^{2}}{2m_{r}}-{\eavehf}\Bigr)^{2}\delta\Bigl(\omega_G^{\rm GOR}-\omega_{\rm BP}-\chi \frac{k^{2}}{2m_{r}}\Bigr)\notag\\
&=\frac{U_{bf}}{4\pi}\left(\frac{2m_{r}U_{bf}N}{\chi}\right)^{\frac{3}{2}}\sqrt{1+\chi\frac{\eavehf}{U_{bf}N}} \,\left[n_f(\xi_{\tilde{k}}^f)+n_b(\xi_{\tilde{k}}^b)\right],
\end{align}
where $\tilde{k}=\sqrt{2m_{r}(\omega_{G}^{\rm GOR}-\omega_{\rm BP})/\chi}=\sqrt{2m_{r}\eavehf+2m_{r}U_{bf}N/\chi}$.
We see that the factor $\chi^{-3/2}$ appears in contrast to the previous order estimate $\tilde{\Phi}\sim \chi^2$.
If the $\omega_{G}^{\rm GOR}$ is far from $\omega_{\rm BP}$, more precisely, if $(\omega_{G}^{\rm GOR}-\omega_{\rm BP})\gg \chi T$, 
the imaginary part is exponentially small by the factor  $\exp[-(\omega_{G}^{\rm GOR}-\omega_{\rm BP})/(\chi T)]$, so the order estimate $\tilde{\Phi}\sim \chi^2$ is still valid.

Since the mass-imbalance effect is generally unavoidable in actual cold atom experiments, in the following we therefore focus on the mass-imbalanced effect on the gap by taking $\mu_f=\mu_b$ and $U_{bf}=U_{bb}$, unless otherwise specified.
As realistic candidates, we consider $^6$Li-$^7$Li, $^{40}$K-$^{41}$K, and $^{173}$Yb-$^{174}$Yb mixtures.
Even in these systems, it is generally difficult to control $U_{bf}$ and $U_{bb}$ independently.
However, in the case of $^6$Li-$^7$Li and $^{40}$K-$^{41}$K mixtures, the boson-boson scattering length $a_{bb}=(m_bU_{bb})/(4\pi)$ can be tuned due to the magnetic Feshbach resonance~\cite{Pollack,Kishimoto}, while the boson-fermion one $a_{bf}=(m_rU_{bf})/(4\pi)$ is almost independent of the magnetic field (noting that $a_{bf}=2.16$ nm~\cite{ENS2017} and $a_{bf}=5.13$ nm~\cite{Falke} in $^6$Li-$^7$Li and $^{40}$K-$^{41}$K mixtures, respectively).
In $^{173}$Yb-$^{174}$Yb mixtures, two scattering lengths are precisely determined as $a_{bf}=7.34$ nm and $a_{bb}=5.55$ nm~\cite{Kitagawa}.

\section{Numerical Results}
\label{sec3}
\begin{figure}[t]
\begin{center}
\includegraphics[width=12cm]{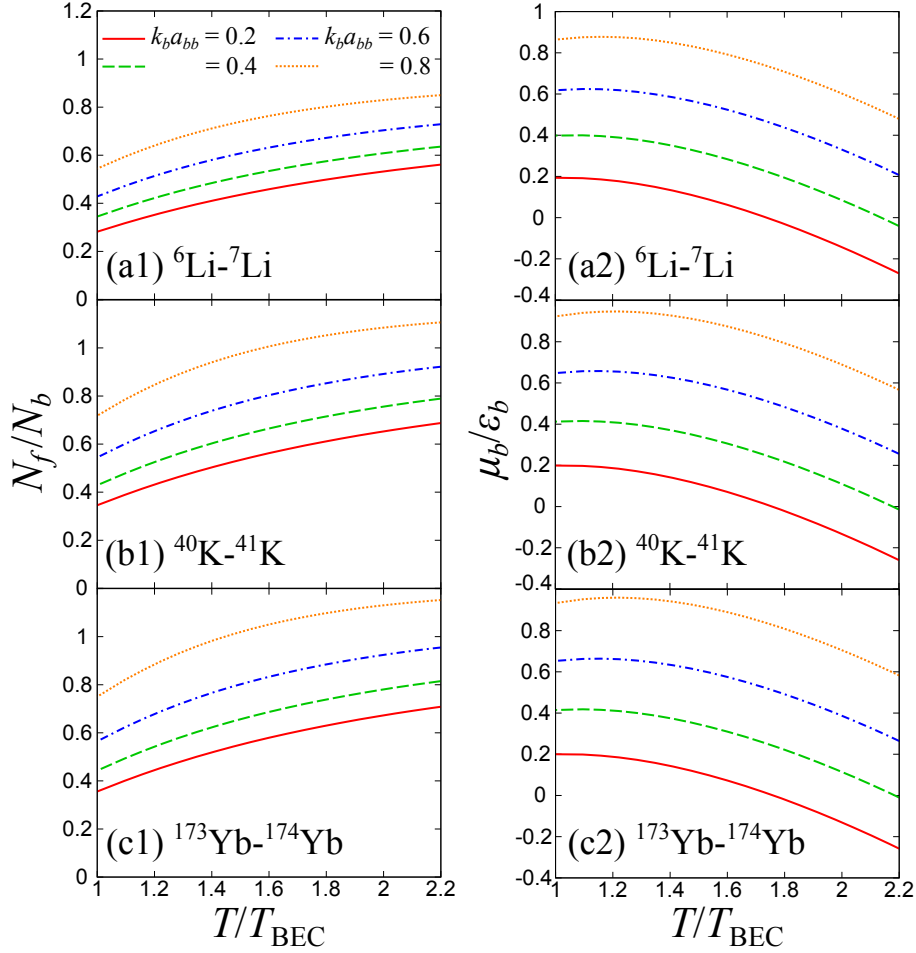}
\end{center}
\caption{Ratio between fermionic and bosonic number densities $N_f/N_b$ with different interactions as functions of the temperature $T$ in (a1) $^6$Li-$^7$Li, (b1) $^{40}$K-$^{41}$K, and (c1) $^{173}$Yb-$^{174}$Yb Bose-Fermi mixtures with $\Delta \mu=\Delta U =0$.
(a2), (b2), and (c2) shows the bosonic chemical potential $\mu_b$ in the systems corresponding to (a1), (b1), and (c1), respectively.
In these plots, $\mu_b$ is divided by the energy scale $\varepsilon_b$ characterizing $N_b$ as {$\varepsilon_b=k_b^2/(2m_b)$ where $k_b=(6\pi^2N_b)^{\frac{1}{3}}$.}
$T_{\rm BEC}$ is the BEC temperature.
}
\label{fig2}
\end{figure}
\subsection{Thermodynamic quantities}

First, we discuss when the system explicitly breaks the supersymmetry
with respect to only the mass-imbalance, namely $U_{bf}=U_{bb}$ and $\mu_f=\mu_b$ but $m_b\neq m_f$.
Figure~\ref{fig2} shows the fermionic number density $N_f$ and the chemical potential $\mu_b=\mu_f$ for three cases with fixed bosonic number density $N_b$, {{at which the two conditions above are realized.}}
The Bose-Einstein condensation temperature $T_{\rm BEC}$ is identified by the Hugenholtz-Pines relation~\cite{HP}
\begin{equation}
\label{eq27}
\mu_b-\Sigma_b=\mu_b-2U_{bb}N_{b}(T=T_{\rm BEC})-U_{bf}N_f(T=T_{\rm BEC})=0.
\end{equation}
{{We see that $N_f$ is smaller than $N_b$ at low $T$.}}
The qualitative temperature dependence of these quantities is unchanged among $^6$Li-$^7$Li, $^{40}$K-$^{41}$K, and $^{173}$Yb-$^{174}$Yb mixtures.

{{This behavior can be understood as follows:}}
In the non-interacting case, $\mu_b=0$ at $T_{\rm BEC}$ and $\mu_b$ is negative above $T_{\rm BEC}$.
In the presence of the interactions, {{the chemical potential is effectively shifted to $\bar{\mu}_{f(b)} \equiv \mu_{f(b)}-\Sigma_{f(b)}^{\rm H}$ due to the Hartree shift.
As $\bar{\mu}_{b}$ is fixed from $N_b$ and $T$, which is negative, $\mu_b$ becomes larger as the interaction strength increases, and eventually becomes positive.}}
On the other hand, $\mu_f$ is positive in the low-temperature regime even in the absence of repulsions due to the Fermi-Dirac statistics.
Therefore, at a weak coupling case, $\mu_f=\mu_b$ would take a positive and small value.
As $N_f$ is an increasing function of $\bar{\mu}_{f}$, which is proportional to $\mu_f$, $N_f$ needs to be much smaller than $N_b$, in order to realize $\mu_f=\mu_b$.
This situation is similar to the so-called Bose polarons~\cite{Santamore,Cucchietti,Rath,Li,JILA,Mistakidis,Takahashi} where impurity atoms (which {{corresponds to}} a fermion in the present case) are immersed a the bosonic medium.
If we increase the interaction, $N_f$ becomes larger and finally {exceeds} $N_b$.
\par
We note that at stronger coupling, the system may be unstable against the phase separation~\cite{Viverit}.
At $T=0$, the mixture is expected to become unstable when $k_ba_{bb}\leq \pi\frac{m_b}{m_b+m_f}\left(\frac{N_b}{N_f}\right)^{\frac{1}{3}}$ at $U_{bb}=U_{bf}$ in Ref.~\cite{Viverit}. 
Since the parameter regimes we consider in this paper are $0.3\lesssim N_f/N_b\lesssim 1.2$ (see Fig.~\ref{fig2}) and $\frac{7}{13}\leq\frac{m_b}{m_f+m_b}\leq\frac{174}{347}$, this stability condition can be estimated as $k_ba_{bb}\lesssim 1.6$.
Furthermore, as usual, such an instability is weakened at finite temperature due to thermal fluctuations.
Therefore, although we do not explicitly address this condition, we assume that the homogeneous phase is realized.
\par
\begin{figure}
\begin{center}
\includegraphics[width=7.5cm]{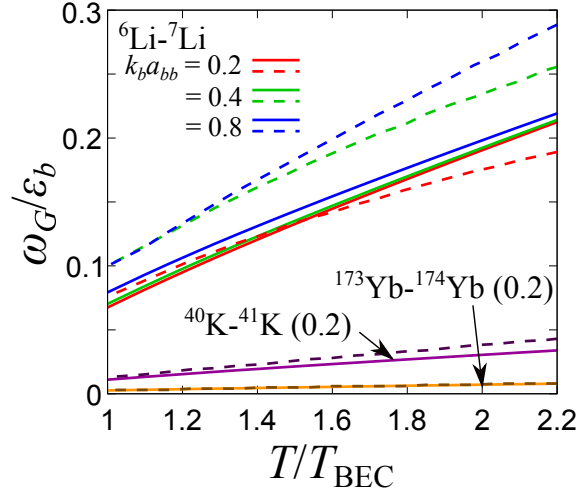}
\end{center}
\caption{Goldstino gap $\omega_G$ calculated by GOR (solid lines) and RPA (dashed lines) with $\Delta U =\Delta\mu=0$.
For $^6$Li-$^7$Li mixtures, we plot $\omega_G$ at different interaction strengths $k_{b}a_{bb}=0.2$, $0.4$, and $0.8$.
We also show $\omega_G$ in $^{40}$K-$^{41}$K and $^{173}$Yb-$^{174}$Yb mixtures at $k_ba_{bb}=0.2$.
}
\label{fig3}
\end{figure}
\subsection{Spectral properties of Goldstino}
Using the thermodynamic quantities shown in Fig.~\ref{fig2} and the GOR relation given by Eq.~(\ref{eq21}) and RPA equation (\ref{eq23}),
we calculate the goldstino gap above $T_{\rm BEC}$ as shown in Fig.~\ref{fig3}.
In the RPA calculation, we have defined $\omega_G^{\rm RPA}$ as the energy where the Goldstino spectral weight $A_G(\bm{p}=\bm{0},\omega)$ has a maximum,
where 
\begin{equation}
\label{eq28}
A_{\rm G}(\bm{p},\omega)={\rm Im}\Gamma^{\rm R}(\bm{p},\omega),
\end{equation}
{{so that it can be defined in the case that the pole has a finite imaginary part.}}
In the case of $^{40}$K-$^{41}$K and $^{173}$Yb-$^{174}$Yb mixtures at $k_ba_{bb}=0.2$,
one can find that the GOR predictions show good agreement with the RPA calculation.
In fact, one can check that the values of expansion parameter estimated in Eq.~\eqref{eq:orderEstimate} are small: $|\omega_G^{\rm GOR}/\omega_{\rm BP}|=9.6\times10^{-2}$ and $|\omega_G^{\rm GOR}/\omega_{\rm BP}|=2.2\times10^{-2}$ in $^{40}$K-$^{41}$K and $^{173}$Yb-$^{174}$Yb mixtures, respectively, at $T=T_{\rm BEC}$.
Also, we see that $\omega_G^{\rm GOR}/\varepsilon_b$ is quite small.
This behavior can be qualitatively understood in the following way: 
The second term in Eq.~(\ref{eq21}) gives the term which is proportional to $\chi \eave_{\rm HF} /\varepsilon_b$, in $\omega_G^{\rm GOR}/\varepsilon_b$.
Assuming that the factor $\eave_{\rm HF} /\varepsilon_b$ is not far from unity, one can make an order estimate of $\omega_G^{\rm GOR}/\varepsilon_b$ by checking  $\chi$. 
Indeed, the values $\chi={1.2}\times10^{-2}$ and $\chi={2.9}\times10^{-3}$ in $^{40}$K-$^{41}$K and $^{173}$Yb-$^{174}$Yb mixtures explain the order of magnitude for $\omega_G^{\rm GOR}/\varepsilon_b$.

Although we do not show the numerical results explicitly at stronger supersymmetric couplings,
this agreement is unchanged in these mixtures.
In this regard, we conclude that the mass-imbalance effect in these systems is negligibly small in this temperature region.
To confirm the existence of Goldstino, exploring the interaction and density dependences of the Goldstino gap given by Eq.~(\ref{eq22}) is suitable.
We note that thermodynamic quantities such as chemical potential can precisely be observed within a relative error of less than 4\% in a recent cold atom experiment~\cite{Horikoshi,TajimaHorikoshi,Horikoshi2}. 
\par 
On the other hand, the mass-imbalance effect on the Goldstino gap in $^6$Li-$^7$Li mixtures with $\chi={1}/13$ is not so small compared to the other two systems. 
In fact, the values of the expansion parameter $|\omega_G^{\rm GOR}/\omega_{\rm BP}|=0.621$, $0.308$ and $0.151$ for 
$k_ba_{bb}=0.2$, $0.4$ and $0.8$ at $T=T_{\rm BEC}$ are not small compared with those in $^{40}$K-$^{41}$K and $^{173}$Yb-$^{174}$Yb mixtures.
We note that the difference between GOR and RPA is accidentally small at $k_ba_{bb}=0.2$. 
This is just a coincidence caused by the singular behavior of the branch point.

\par
\begin{figure}[t]
\begin{center}
\includegraphics[width=8cm]{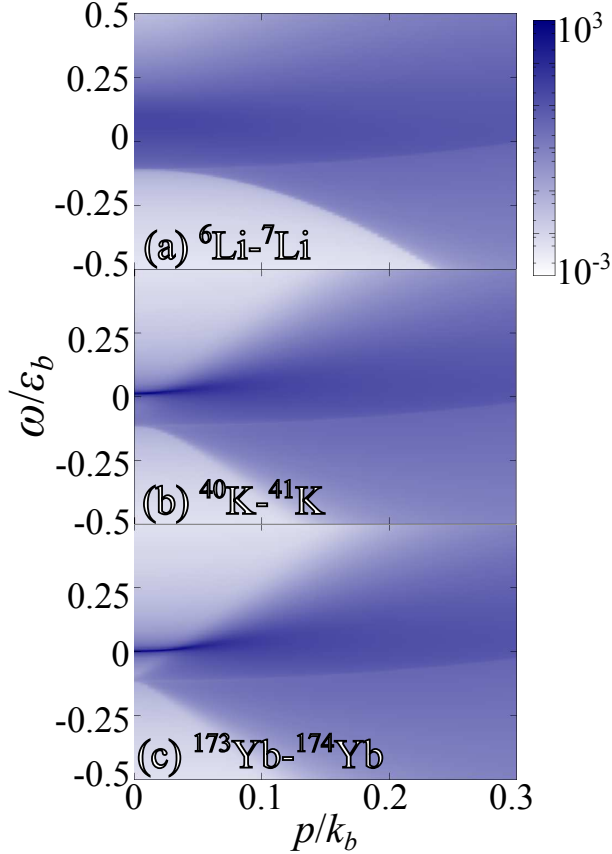}
\end{center}
\caption{Calculated Goldstino spectral weight $A_G(\bm{p},\omega)\varepsilon_b/N_b$ within RPA in (a) $^6$Li-$^7$Li, (b) $^{40}$K-$^{41}$K, and (c) $^{173}$Yb-$^{174}$Yb Bose-Fermi mixtures at $T=T_{\rm BEC}$.
The parameters are set at $k_ba_{bb}=0.2$ and $\Delta U=\Delta \mu =0$.
}
\label{fig4}
\end{figure}
Figure~\ref{fig4} shows the RPA spectral weight $A_{\rm G}(\bm{p},\omega)$ of the Goldstino at $T=T_{\rm BEC}$, $k_{b}a_{bb}=0.2$, {{and finite momentum}}.
While $A_{\rm G}(\bm{p},\omega)$ in 
$^{40}$K-$^{41}$K and $^{173}$Yb-$^{174}$Yb mixtures
exhibit a sharp peak associated with the supersymmetric collective mode around $\omega=0$,
such a peak in a $^6$Li-$^7$Li mixture is strongly broadened due to the branch point at weak coupling.
We note that the Goldstino spectrum merges with the continuum at finite momenta even in $^{40}$K-$^{41}$K and $^{173}$Yb-$^{174}$Yb mixtures.
\par
\begin{figure}[t]
\begin{center}
\includegraphics[width=7.5cm]{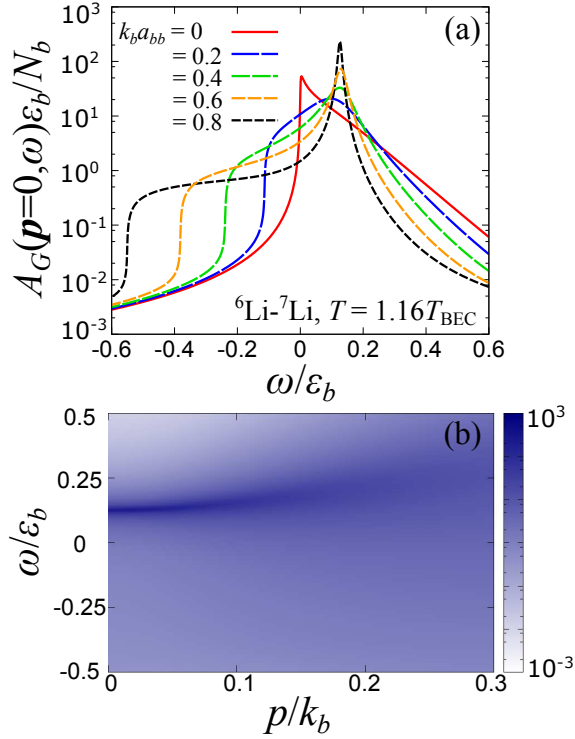}
\end{center}
\caption{(a) Goldstino spectral weight $A_G(\bm{p}=\bm{0},\omega)$ at zero momentum in a $^6$Li-$^7$Li mixture at $T=1.16T_{\rm BEC}$ and $\Delta\mu = \Delta U =0$ with $k_ba_{bb}=0, 0.2, 0.4, 0.6, 0.8$.
(b) Contour plot of $A_G(\bm{p},\omega)$ at $k_{b}a_{bb}=0.8$. 
}
\label{fig5}
\end{figure}
Figure~\ref{fig5}(a) shows $A_G(\bm{p}=\bm{0},\omega)$ at zero momentum in a $^6$Li-$^7$Li mixture at $T=1.16T_{\rm BEC}$.
With increasing the supersymmetric interaction $k_{b}a_{bb}$ ($=\frac{m_b}{\blue{2}m_r}a_{bf}$ since we take $U_{bb}=U_{bf}$),
one can see the crossover from the regime where the singularity associated with $\omega_{\rm BP}$ is dominant, to the coexistence of the sharp Goldstino pole and continuum plateau.
{{It is possible to check the sharp Goldstino peak also at finite momentum, from the spectral function at $k_ba_{bb}=0.8$ plotted in Fig.~\ref{fig5}(b)}}.
While in the non-interacting case ($k_{b}a_{bb}=0$) a kink structure can be found around $\omega=0$, it originates from mainly the {{tip of the continuum, as one can see from Appendix~\ref{AppB}}}.
We note that the contribution at $\omega<0$ in the non-interacting case is {{an artifact}} associated with the small imaginary part $i\delta=10^{-3}i$.
{{We also note that, some of the analysis above}} in a tight-binding model {{were done in Ref.~\cite{Shi}, but this is the first time that we got results for gases of realistic Bose-Fermi mixtures.}}
\par
\begin{figure}[t]
\begin{center}
\includegraphics[width=7.5cm]{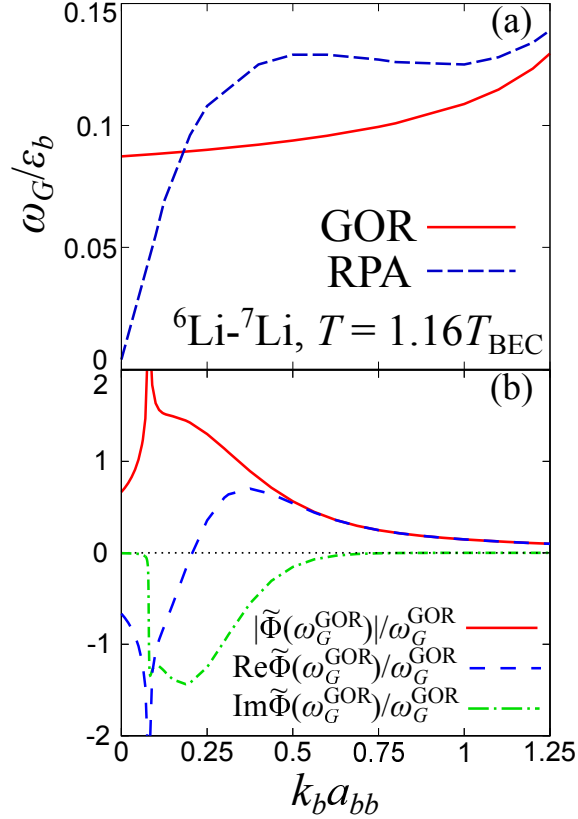}
\end{center}
\caption{(a) Comparison of the Goldstino gap $\omega_G$ between GOR and RPA and (b) $\tilde{\Phi}(\omega_G^{\rm GOR})/\omega_G^{\rm GOR}$ given by Eq.~(\ref{eqphi}) 
in a $^6$Li-$^7$Li mixture at $T=1.16T_{\rm BEC}$ and $\Delta \mu = \Delta U = 0$.
}
\label{fig6}
\end{figure}
Figure~\ref{fig6}(a) shows a comparison of the Goldstino gap from the GOR relation and the RPA calculation (corresponding to Fig~\ref{fig5}) in a $^6$Li-$^7$Li mixture at $T=1.16T_{\rm BEC}$.
If we increase $k_ba_{bb}$ (and simultaneously $k_ba_{bf}$ such that $U_{bf}=U_{bb}$),
one can find that two results approach each other around $k_{b}a_{bb}\gesim 0.6$.
In such a regime, as shown in Fig.~\ref{fig5}, a sharp Goldstino peak appears since the branch point is separated from the pole.
We also plot $\tilde{\Phi}(\omega_G^{\rm GOR})/\omega_G^{\rm GOR}$ given by Eq.~(\ref{eqphi}) in Fig.~\ref{fig6}(b). 
Since $\tilde{\Phi}(\omega_G^{\rm GOR})$ represents the higher-order corrections included in RPA, 
the GOR relation is expected to be valid in the region where $\tilde{\Phi}(\omega_G^{\rm GOR})/\omega_G^{\rm GOR}$ is small. 
In this sense, one can obtain the strong intensity of Goldstino pole even in $^6$Li-$^7$Li mixture in the presence of the relatively strong interactions.
Such a condition in this case can be expressed as $\chi\langle\mathcal{E}\rangle\ll U_{bf}N$.  
From Fig.~\ref{fig6}(b), one can confirm that the coincidence of $\omega_G^{\rm GOR}$ and $\omega_G^{\rm RPA}$ at $k_ba_{bb}\simeq 0.2$ in Figs.~\ref{fig3} and \ref{fig6}(a) is accidental due to ${\rm Re}\tilde{\Phi}(\omega_G^{\rm GOR})=0$.
\par
\begin{figure}[t]
\begin{center}
\includegraphics[width=7.5cm]{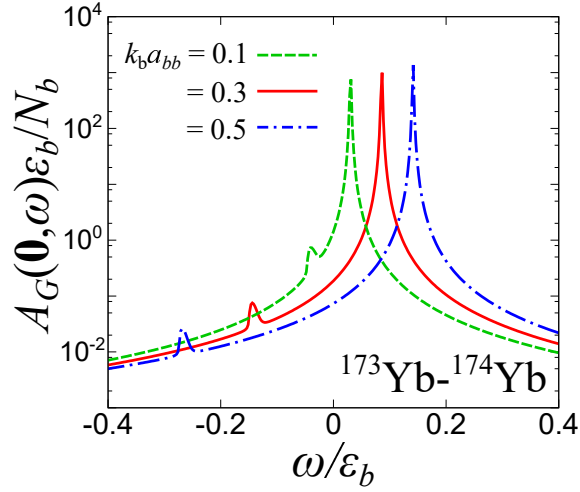}
\end{center}
\caption{Calculated Goldstino spectral weight $A_G(\bm{p}=\bm{0},\omega)\varepsilon_b/N_b$ within RPA in a $^{173}$Yb-$^{174}$Yb Bose-Fermi mixture at $T= T_{\rm BEC}$ and $\Delta\mu=0$.
The interaction parameters are chosen to reproduce the experimental ratio $a_{bf}/a_{bb}=7.34/5.55$~\cite{Kitagawa}.   
The sharp peaks at positive energy are the Goldstino poles.
}
\label{fig7}
\end{figure}

Furthermore, we plot in Fig.~\ref{fig7} $A_G(\bm{0},\omega)$ in a $^{173}$Yb-$^{174}$Yb mixture with realistic interactions given by $a_{bf}/a_{bb}=7.34/5.55$~\cite{Kitagawa}, at $T=T_{\rm BEC}$ and $\mu_f=\mu_b$.
A sharp peak of the Goldstino pole emerges at a positive energy, whereas a small peak of the continuum is in the negative energy region.
The continuum is quite small compared to the case of a $^6$Li-$^7$Li mixture shown in Fig.~\ref{fig5}.
We have checked that the GOR relation shows an excellent agreement with the pole position in this case.
If we increase $N_b$ (namely, the coupling parameter $k_{b}a_{bb}=(6\pi^2N_b)^{\frac{1}{3}}a_{bb}$ with fixed $a_{bb}$),
the Goldstino pole becomes distinct since the continuum goes to the lower-energy region.
This result indicates that the observation of the Goldstino gap in $^{173}$Yb-$^{174}$Yb mixtures is quite promising. 
\section{Fermionic single-particle  spectrum}
\label{sec:fermionSpectrum}
\begin{figure}[t]
\begin{center}
\includegraphics[width=4cm]{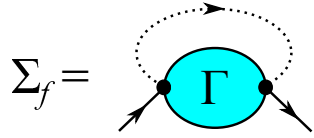}
\end{center}
\caption{Self-energy diagram for supersymmetric fluctuations associated with the RPA Goldstino propagator $\Gamma$.
The solid (dashed) lines represent fermion (boson) propagators $G_{b(f)}^{\rm H}$ with the Hartree shift $\Sigma_{f(b)}^{\rm H}$.
The black dots are the boson-fermion coupling $U_{bf}$.
}
\label{fig8}
\end{figure}

In this section, in addition to Tan's contacts shown in Sec.~\ref{sec2}, we discuss how to detect the Goldstino gap in cold atom experiments.
One of promising ways is the single-particle excitation spectrum of a fermion as discussed in the BEC phase~\cite{Satow2}.
In the normal phase, we consider the self-energy $\Sigma_{f}(\bm{p},i\omega_\ell)$ diagrammatically drawn in Fig.~\ref{fig8}, where $\omega_\ell=(2\ell+1)\pi T$ is the Matsubara frequency for fermions.
The explicit form of  $\Sigma_{f}(\bm{p},i\omega_\ell)$ is given by
\begin{equation}
\label{eq29}
\Sigma_f(\bm{p},i\omega_\ell)=-U_{bf}^2T\sum_{n=-\infty}^{\infty}\int \frac{d^{3}\bm{k}}{(2\pi)^{3}}\Gamma(\bm{k},i\omega_n)G_b^{\rm H}(\bm{p}-\bm{k},i\omega_\ell-i\omega_n),
\end{equation}
where $G_b^{\rm H}(\bm{p},i\tilde{\omega}_k)=1/(i\tilde{\omega}_k-\xi_{\bm{p}}^b)$ is the Hartree Green's function of a boson.
Here, $\tilde{\omega}_k=2k\pi  T$ is the Matsubara frequency for bosons.
We obtain the single-particle Green's function $G_f(\bm{p},i\omega_\ell)$ as
\begin{equation}
\label{eq30}
G_f(\bm{p},i\omega_\ell)=\frac{1}{i\omega_\ell-\xi_{\bm{p}}^f-\Sigma_f(\bm{p},i\omega_\ell)}.
\end{equation}
The fermionic single-particle spectral function is obtained as $A_f(\bm{p},\omega)=-\frac{1}{\pi}{\rm Im}G_f(\bm{p},i\omega_\ell\rightarrow \omega+i\delta)$.
\par
\par
\begin{figure}[t]
\begin{center}
\includegraphics[width=7cm]{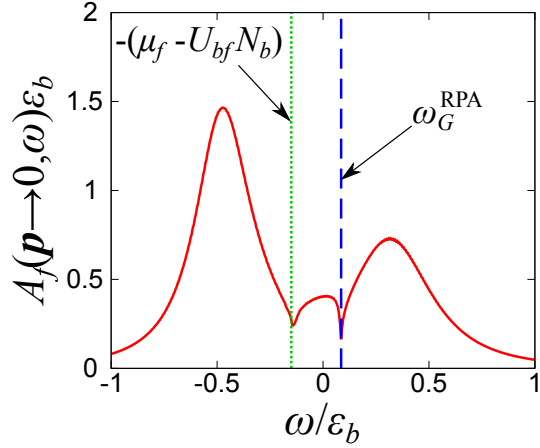}
\end{center}
\caption{{Fermionic single-particle spectral function $A_f(\bm{p}\rightarrow\bm{0},\omega)$ at the zero-momentum limit in a $^{173}$Yb-$^{174}$Yb mixture at $T=T_{\rm BEC}$ with $k_{b}a_{bb}=0.3$ and $a_{bf}/a_{bb}=7.34/5.55$~\cite{Kitagawa}.
In the numerical calculation, we take the momentum cutoff $\Lambda=2k_b$.
The vertical dotted and dashed lines represent the fermionic pole position $\xi_{\bm{p}\rightarrow\bm{0}}^f=-(\mu_f-U_{bf}N_b)$ within the mean-field theory and the Goldstino gap $\omega_{G}^{\rm RPA}$ obtained from the RPA analysis.}
}
\label{fig9}
\end{figure}

{
Figure~\ref{fig9} shows the calculated $A_f(\bm{p}\rightarrow \bm{0},\omega)$ in a $^{173}$Yb-$^{174}$Yb mixture at $T=T_{\rm BEC}$ with realistic interactions, that is, $k_{b}a_{bb}=0.3$ and $a_{bf}/a_{bb}=7.34/5.55$~\cite{Kitagawa}.
One can find that $A_f(\bm{p}\rightarrow \bm{0},\omega)$ contains a double-peak structure due to the self-energy correction.
In the absense of such a correction, 
only the single fermionic pole locates at $\omega=\xi_{\bm{p}\rightarrow\bm{0}}^{f}\equiv-(\mu_f-U_{bf}N_b)$.
As we mentioned in Sec.~\ref{sec2}, in the zero-range model,
the fluctuation effect beyond the mean-field theory involves an ultraviolet divergence.
In this regard, we take a finite momentum cutoff $\Lambda=2k_b$ in Eq.~(\ref{eq29}).
We note that the double-peak structure in $A_f(\bm{p}\rightarrow \bm{0},\omega)$ is left qualitatively unchanged by the value of $\Lambda$.  
}
\par
We examine a qualitative structure of $A_{f}(\bm{p},\omega)$ by focusing on the Goldstino pole and using an approximate form of $\Gamma$ as
\begin{equation}
\label{eq31}
\Gamma(\bm{k},i\omega_n)\simeq\frac{Z_G}{i\omega_n-E_{\bm{k}}},
\end{equation}
where $E_{\bm{k}}=k^2/(2m_G)+\omega_G$ is the Goldstino dispersion.
$Z_{G}$ and $m_G$ are the wave-function renormalization and the effective mass of the Goldstino, respectively.
{{Their analytical expressions are obtained in the supersymmetric case at $T=0$~\cite{Satow2}.}}
{{By using this expression}}, we can analytically perform the summation of fermion Matsubara frequency in $\Sigma_{f}(\bm{p},i\omega_\ell)$
as
\begin{equation}
\label{eq32}
\Sigma_f(\bm{p},i\omega_\ell)=U_{bf}^2Z_G\int \frac{d^{3}\bm{k}}{(2\pi)^{3}}\frac{1-n_f(E_{\bm{p}-\bm{k}})+n_b(\xi_{\bm{k}}^b)}{i\omega_\ell-E_{\bm{p}-\bm{k}}-\xi_{\bm{k}}^b}.
\end{equation}
Furthermore, near $T=T_{\rm BEC}$ {{it was suggested that}} one can use the so-called static approximation where $n_{b}(\xi_{\bm{k}}^b)$ has a dominant contribution at $\xi_{\bm{k}}^b=0$~\cite{Kharga}.
{{For qualitative illustration purpose, we use this approximation and obtain}} 
\begin{equation}
\label{eq33}
\Sigma_f(\bm{p},i\omega_\ell)\simeq\frac{U_{bf}^2Z_GN_b}{i\omega_\ell-E_{\bm{p}}}.
\end{equation}
Finally, the fermionic spectral function $A_f(\bm{p}\rightarrow\bm{0},\omega)$ at the zero-momentum limit reads
\begin{equation}
\label{eq34}
A_f(\bm{p}\rightarrow\bm{0},\omega)=\alpha_{+}\delta(\omega-E_{+})+\alpha_{-}\delta(\omega-E_{-}),
\end{equation}
where
\begin{equation}
\label{eq35}
E_{\pm}=\frac{\omega_G-\mu_f+U_{bf}N_b}{2}\pm\sqrt{\left(\frac{\omega_G+\mu_f-U_{bf}N_b}{2}\right)^2+U_{bf}^2Z_GN_b},
\end{equation}
and
\begin{equation}
\alpha_{\pm}=\frac{1}{2}\left(1 {{\mp}} \frac{\omega_G+\mu_f-U_{bf}N_b}{\sqrt{(\omega_G+\mu_f-U_{bf}N_b)^2+4U_{bf}^2Z_GN_b}}\right).
\end{equation}
{{This double-peak structure is due to the level repulsion between the one-particle fermion excitation and the Goldstino pole in $A(\bm{p}\rightarrow\bm{0},\omega)$.}}
{This level repulsion enlarges the separation between the fermionic pole $-(\mu_f-U_{bf}N_b)$ and the Goldstino pole $\omega_{G}^{\rm RPA}$.
}
One can estimate $\omega_G$ from {{$E_\pm$ and $\alpha_\pm$}}.
Indeed, in cold atom experiments, radio-frequency spectroscopies are employed to observe single-particle excitations~\cite{Torma}.
If the interaction and chemical potential dependences of  the low-momentum excitation spectra are observed, one can estimate $\omega_G$ from them.
\section{Summary}
\label{sec4}
To summarize, we have theoretically investigated the gapped Goldstino mode in an ultracold atomic Bose-Fermi mixture with explicitly broken supersymmetry.
We have shown the gap formula for the Goldstino (GOR relation) by using the memory function formalism.
Using this relation, we calculate the Goldstino gap at the first order of the explicit symmetry breaking, and compare it with the numerical results obtained within the RPA.
We have confirmed that in the absence of the mass-imbalance between fermions and bosons, the Goldstino gap obtained by the GOR relation coincides with that in the RPA.
We have also discussed the relationship between the derived GOR relation and Tan's contact.
At the current stage, a $^{173}$Yb-$^{174}$Yb mixture is the strongest candidate to detect the Goldstino. Indeed, using experimental values of scattering lengths and mass-ratio, we show that the Goldstino pole has a strong intensity in this mixture.
While the mass-imbalance effect in $^{40}$K-$^{41}$K and $^{173}$Yb-$^{174}$Yb mixtures
is negligibly small even in the weak-coupling regime,
that in a $^6$Li-$^7$Li mixture induces broadening of the Goldstino pole due to the singularity around the branch point in fermion-boson bubbles {{through the infinite sum of bubbles in the RPA}}.
However, if we increase the interactions, the Goldstino pole becomes sharp even in this case since the branch point is well separated from the Goldstino pole.
Finally, we have discussed a possibility of observing the Goldstino gap from the single-particle excitation of the Fermi atom.
We show a qualitative structure of the spectral function near $T=T_{\rm BEC}$ and predict the modification of the dispersion due to the coupling between the free branch and the Goldstino pole at low-momenta.
\par
To further address realistic experimental situations,
it is an important problem to investigate the radio-frequency spectra and the momentum-resolved photoemission spectra of the Fermi atom in the present mixtures.
In such a case, we have to consider the inhomogeneity due to the trap potential.
In addition, exploring other approaches to see the Goldstino gap, such as nonequilibrium dynamics~\cite{Danshita,Siegl}, are also interesting future directions.

\acknowledgments
We would like to thank M. Horikoshi for useful discussion.
H.T. was supported by Grant-in-Aid for JSPS fellows (No.17J03975) and for Scientific Research from JSPS (No.18H05406).
Y.H. is supported in part by Japan Society of Promotion of Science (JSPS) Grant-in-Aid for Scientific Research
(KAKENHI Grants No. 16K17716, 17H06462, and 18H01211).

\appendix

{
\section{Memory function formalism}\label{sec:MemoryFunction}
In this Appendix, we show the detailed derivation of Eq.~\eqref{eq:retardedGreenFunction} based on the memory function formalism~\cite{Mazenko}.
This formalism is useful to describe the Langevin dynamics of slow variables such as hydrodynamic degrees of freedom.
We would like to obtain the equation of $\Gamma^R(\omega)$ with the following form:
\begin{equation}
\label{eq:GREquation}
(\omega+\Omega +i\Phi(\omega))\Gamma^R(\omega)= -N,
\end{equation}
where $\Omega$ and $N$ are constants, and $\Phi(\omega)$ is a function of $\omega$. 
In the language of the memory function formalism, $K(\omega)  =  i \Omega -\Phi(\omega)$ is called the memory function;  $i \Omega$ and $\Phi(\omega)$ are static and dynamical parts of the memory function.
The corresponding generalized Langevin equation reads
\begin{equation}
\begin{split}
\label{eq:Langevin}
(\partial_{t}-i\Omega)Q(t) + \int^{t}_{0} dt'\Phi(t-t')Q(t') = R(t),
\end{split}
\end{equation}
where we introduce the noise $R(t)$ that satisfies $\langle Q(t)R(t') \rangle=0$. We do not give the explicit relation between Eqs.~\eqref{eq:GREquation} and \eqref{eq:Langevin}, 
which can be shown by using the projection operator method~\cite{Mori,Zwanzig}.
Equation~\eqref{eq:GREquation} has a similar form to the Schwinger-Dyson equation in a quantum field theory.
Roughly speaking, $\Omega +i\Phi(\omega)$ corresponds to the self-energy, $\Omega$ and $\Phi$ give a gap and dissipation, respectively.
The purpose here is to express $\Omega$, $N$, and $\Phi(\omega)$ by correlation functions.
For this purpose, we introduce the Liouville operator $\mathcal{L}$ as $\mathcal{L}q\equiv[H,q]$ such that we can express $q(\bm{r},t)$ as $q(\bm{r},t)=e^{i\mathcal{L}t}q(\bm{r},0)$. This enable us to rewrite Eq.~(\ref{eq5}) at $\bm{p}=\bm{0}$ as
\begin{equation}
\label{eq7}
\Gamma^{R}(\omega)=i\langle\{r(\omega)Q,q^\dag(\bm{0},0)\}\rangle,
\end{equation}
where $r(\omega)=-i/(\omega+\mathcal{L})$.
Using the identity  $-i\omega r(\omega)=1+i\mathcal{L}r(\omega)$, we obtain
\begin{equation}
\label{eq8}
-i\omega\Gamma^R(\bm{0},\omega)=iN+i\langle\{r(\omega)i\mathcal{L}Q,q^\dag(\bm{0},0)\}\rangle,
\end{equation}
where
\begin{equation}
\label{eq:TotalNumberDensity}
\begin{split}
N=\langle\psi_{f}^{\dag}(\bm{r})\psi_{f}(\bm{r})\rangle+\langle\psi_{b}^{\dag}(\bm{r})\psi_{b}(\bm{r})\rangle
\end{split}
\end{equation}
is the total number density.
We now introduce the memory function $K(z)$ such that
\begin{equation}
\label{eq9}
K(\omega)\Gamma^R(\bm{0},\omega)=i\langle\{r(\omega)i\mathcal{L}Q,q^\dag(\bm{0},0)\}\rangle.
\end{equation}
By construction, it satisfies $(\omega -iK(\omega))\Gamma^{R}(\omega)=-N$.
We would further like to decompose $K(\omega)$ into the static part $i\Omega$ that is responsible for the gap
and the dynamic one $\Phi(\omega)$ that is responsible for the dissipation.
Multiplying  Eq.~\eqref{eq9} by $-i\omega$ and using Eq.~\eqref{eq8} and $-i\omega r(\omega)=1+i\mathcal{L}r(\omega)$, we obtain
\begin{align}
\label{eq:Kz1}
&K(\omega)
\left[
i N +  i  \langle \{r(\omega) i \mathcal{L} Q,q^{\dag}(\bm{0},0)\}\rangle 
\right]\notag\\
&\quad  =   i  \langle \{i \mathcal{L}Q,q^{\dag}(\bm{0},0)\}\rangle
+i  \langle \{r(\omega)(i \mathcal{L})^{2}Q,q^{\dag}(\bm{0},0)\}\rangle.
\end{align}
From Eq.~\eqref{eq9}, the left hand side of Eq.~\eqref{eq:Kz1} can be written as
\begin{equation}
\label{eq:Kz2}
i N K(\omega) +  
i\frac{i}{\Gamma^R(\bm{0},\omega)} \langle \{r(z) i \mathcal{L}Q,q^{\dag}(\bm{0},0)\}\rangle^2 .
\end{equation}
Substituting Eq.~\eqref{eq:Kz2} into Eq.~\eqref{eq:Kz1},
we obtain $K(\omega)  =  i \Omega -\Phi(\omega)$ with
\begin{align}
i \Omega &=  \frac{1}{ N }  \langle \{i \mathcal{L}Q,q^{\dag}(\bm{0},0)\}\rangle, \label{eq:Omega}\\
\Phi(\omega) &= -\frac{1}{ N }  \langle \{r(\omega) (i \mathcal{L})^{2}Q,q^{\dag}(\bm{0},0)\}\rangle 
+\frac{1}{ N } 
\frac{i}{\Gamma^R(\bm{0},\omega)}
  \langle \{ r(\omega)i \mathcal{L}Q,q^{\dag}(\bm{0},0)\}\rangle^2 .
\end{align}
Noting the following relation:
\begin{equation}
\langle \{i\mathcal{L}A,B\} \rangle =\frac{i}{\mathrm{tr} e^{-\beta H}}\mathrm{tr} e^{-\beta H}\{[H,A],B\}
=\frac{-i}{\mathrm{tr} e^{-\beta H}}\mathrm{tr} e^{-\beta H}\{A,[H,B]\}
 =  -\langle \{A,i\mathcal{L}B\} \rangle,
\end{equation}
we can express the dynamic part as 
\begin{align}\label{eq:Phiz}
\Phi(\omega) &= \frac{1}{ N }  \langle \{r(\omega) i \mathcal{L}Q,i \mathcal{L}q^{\dag}(\bm{0},0)\}\rangle\notag\\
&\quad
+\frac{1}{ N }  \langle \{ r(\omega)i \mathcal{L}Q,q^{\dag}(\bm{0},0)\}\rangle
\frac{i}{\Gamma^R(\bm{0},\omega)}
  \langle \{ r(\omega)Q,i \mathcal{L}q^{\dag}(\bm{0},0)\}\rangle
 . 
\end{align} 
In summary, the retarded Green function satisfies Eq.~\eqref{eq:GREquation}, which corresponds to the generalized Langevin equation~\eqref{eq:Langevin}.
The coefficients and function are given by Eqs.~\eqref{eq:TotalNumberDensity}, \eqref{eq:Omega}, and \eqref{eq:Phiz}.

Let us check the order of $\Omega$ and $\Phi(\omega)$ with respect to the explicit breaking term.
Since both static and dynamic parts are proportional to $i \mathcal{L}Q = i [H, Q]$, $K(\omega)$ vanishes if the supersymmetry is exact.
Therefore, $\omega=0$ becomes the pole.
When the supersymmetry is explicitly broken by a small parameter, $[H, Q]\sim \epsilon$,
the static part is $i \Omega\sim \epsilon$, while the dynamic part is $\Phi(\omega) \sim \epsilon^{2}$ as shown in Eq.~\eqref{eq:Phiz}.
Therefore, at the leading order in $\epsilon$, we can neglect $\Phi(\omega)$. 
We note that we also assigned $[H, q^{\dag}(\bm{x},t)]\sim \epsilon$. Precisely speaking, there is the other contribution of order one, the divergence of supersymmetric current $\nabla\cdot \bm{j}$ in $[H, q^{\dag}(\bm{x},t)]$. This vanishes in the correlation function at $\bm{p}=\bm{0}$.
At the leading order in $\epsilon$,  we find the pole~\eqref{eq12}.
}

\section{Analytical results in a non-interacting mixture}
\label{AppB}
{{To check that, the numerical procedure to include infinitesimal imaginary part $i\delta$ in the analytic continuation, does not cause serious numerical artifact, we investigate the Goldstino spectral function in the free limit.}}
In the non-interacting case, the spectral weight $A_{G,0}(\bm{p}=\bm{0},\omega)$ reads
\begin{align}
\label{eqb1}
A_{G,0}(\bm{0},\omega)
&=-{\rm Im}\int\frac{d^3\bm{k}}{(2\pi)^3}\frac{n_{b}(\xi_{\bm{k}}^b)+n_f(\xi_{\bm{k}}^f)}{\omega+\Delta\mu+i\delta-\chi k^2/2}\cr
&=\int_0^{\infty}\frac{k^2dk}{2\pi}\left[n_{b}(\xi_{\bm{k}}^b)+n_f(\xi_{\bm{k}}^f)\right]
\delta\left(\omega+\Delta\mu-\chi \frac{k^2}{2}\right) \quad (\delta\rightarrow 0).
\end{align}
By performing the momentum integration,
we can obtain an analytical expression of $A_G(\bm{0},\omega)$ as
\begin{align}
\label{eqb2}
A_{G,0}(\bm{0},\omega)&=\frac{1}{\sqrt{2}\pi\chi^{\frac{3}{2}}}\theta(\omega)
\sqrt{\omega}
\left[\frac{1}{e^{\left(\frac{\omega}{m_b\chi}-\mu_b\right)/T}-1}
+\frac{1}{e^{\left(\frac{\omega}{m_f\chi}-\mu_f\right)/T}+1}
\right],
\end{align}
{{at $\Delta\mu=0$.}}
If we take $\chi\rightarrow 0$, $A_{G,0}(\bm{0},\omega)$ diverges at only $\omega=0$, {{which indicates that the continuum has vanishing width at zero momentum.}}
{{It can be understood also from the sum rule~\cite{Satow1,Satow2}, $\int d\omega A_{G}(\bm{p},\omega)/\pi= N$:
At finite $\omega$, $A_{G,0}$ vanishes at $\chi\rightarrow 0$, due to the exponential factor.
Therefore, to satisfy the sum rule, existence of divergence at $\omega=0$ is implied.
}}

\begin{figure}[t]
\begin{center}
\includegraphics[width=7.5cm]{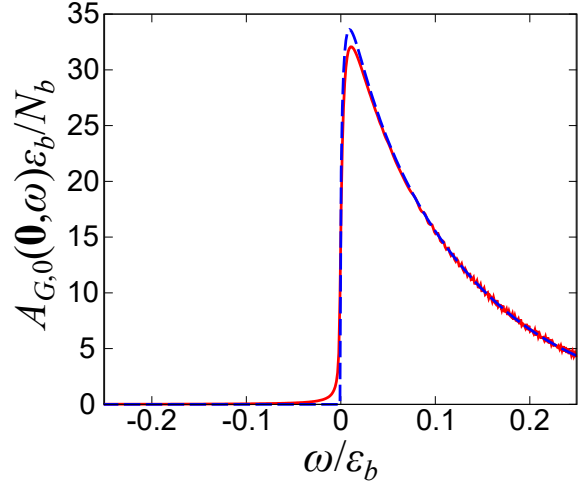}
\end{center}
\caption{Comparison of $A_{G,0}(\bm{0},\omega)$ in a non-interacting $^6$Li-$^7$Li mixture at $T=1.38T_{\rm BEC}$ with $\mu_b=\mu_f$, obtained from Eq.~(\ref{eqb2}) (dashed curve) and the numerical result with $\delta=10^{-3}$ (solid curve).
}
\label{figb1}
\end{figure}

In Fig.~\ref{figb1} we show a comparison between Eq.~(\ref{eqb2}) and the numerical result with $\delta=10^{-3}$, in a non-interacting $^6$Li-$^7$Li mixture at $T=1.38T_{\rm BEC}$.
From this, we can find that effects of $\delta$ is small for the maximum of $A_{G,0}(\bm{0},\omega)$.
While the finite contribution at $\omega<0$ in the numerical calculation originates from finite $\delta$,
we confirmed that this also does not affect our main results.

\end{document}